\documentclass[aps, showpacs, showkeys,nofootinbib,floatfix]{revtex4}

\usepackage{amssymb}
\usepackage{amsmath}
\usepackage{graphicx}
\usepackage{hyperref}
\usepackage{graphicx}

\begin{document}

\title{Scattering of scalar perturbations with cosmological constant in low-energy and high-energy regimes}
\author{Molin Liu$^{1}$}
\email{mlliu@mail2.xytc.edu.cn}
\author{Benhai Yu$^{1}$}
\author{Rumin Wang$^{1}$}
\author{Lixin Xu$^{2}$}
\affiliation{$^{1}$College of Physics and Electronic Engineering,
Xinyang Normal University, Xinyang, 464000, P. R. China\\
$^{2}$School of Physics and Optoelectronic Technology,
Dalian University of Technology, Dalian, 116024, P. R. China}

\begin{abstract}
We study the absorption and scattering of massless scalar waves propagating in spherically symmetric spacetimes with dynamical cosmological constant both in low-energy and high-energy zones. In the former low-energy regime, we solve analytically the Regge-Wheeler wave equation and obtain an analytic absorption probability expression which varies with $M\sqrt{\Lambda}$, where $M$ is the central mass and $\Lambda$ is cosmological constant. The low-energy absorption probability, which is in the range of $[0, 0.986701]$, increases monotonically with increase in $\Lambda$. In the latter high-energy regime, the scalar particles adopt their geometric optics limit value. The trajectory equation with effective potential emerges and the analytic high-energy greybody factor, which is relevant with the area of classically accessible regime, also increases monotonically with increase in $\Lambda$, as long $\Lambda$ is less than or of the order of $10^4$. In this high-energy case, the null cosmological constant result reduces to the Schwarzschild value $27\pi r_g^2/4$.
\end{abstract}



\maketitle

\section{Introduction}
The absorption and scattering of particles outside
a black hole has been a topic of research since 1970s. This issue refers to many key theoretical physics problems, such as quantum entanglement\cite{quanentang,quanentang11,quanentang22}, entropy\cite{entropybekenstein,entropybekenstein11}, information \cite{information,information11,information22,information33} which arise in black hole physics and quantum gravity where fundamental issues about gravity physics and the structure of space are of interest. After Hawking \cite{Hawking,Hawking11,Hawking22} proved  black hole can emit thermal radiation via quantum field theory, the absorption and scattering issue is focused gradually in black hole physics. Many people, such as Unruh \cite{Unruh}, S$\acute{a}$nchez\cite{Sanchez1,Sanchez2,Sanchez3}, Das \cite{Das}, Emparan\cite{Emparan,Emparan11} etc., have done a lot of works on the black hole scattering in the aspects of calculation of absorption greybody factor and cross section as well as black hole's based properties such as unitarity and reciprocity \cite{Sanchez1}, phase shifts and diffraction pattern \cite{Sanchez2}, angular distribution \cite{Sanchez3}, extra dimensional effect \cite{Das,Emparan,Emparan11} and so on. What this paper \cite{Unruh} refers to is the Unruh's work in which the original analytical method to calculating black hole cross section is shown in detail in 1976. The radiation equation, i.e. Regge-Wheeler equation in perturbation theory, is broken down into three sub-equations in following three regions: one is near the horizon, one is intermediate region and the last is far from the black hole. Then these sub-equations can be solved easily by means of well-known functions or specific functions. Finally, the absorption coefficients are obtained by matching these solutions in the areas of overlap between the regions. It is found that the absorption cross section of Dirac particles is exactly 1/8 of the massive scalar particles. The total cross section of massive scalar field is a function of ingoing particles frequency. Scalar cross section $\sigma_s$ and transmission coefficients $(T_s)_l$ \cite{Unruh} are
\begin{eqnarray}
  \sigma_s &=& \frac{(4\pi M)^2 (1 + v^2) (2 M m)}{v^2 (1 - v^2)^{1/2}\{1- \exp \left[-2 \pi M m (1 + v^2)/v (1 - v^2)^{1/2}\right]\}}, \label{crosec}\\
  (T_s)_l &=& \frac{\pi l!^4 2^{2l+2}\omega (1+v^2) \omega^{2l+2} v^{2l}}{(2l)!^4 (2l+1)^2\{1 - \exp\left[-\pi \omega (1+v^2)/v\right]\}} \prod_{s=1}^{l}\left[s^2 +\left(\frac{\omega (1 + v^2)}{2v}\right)^2\right], \label{trancoe1}
\end{eqnarray}
where $v = \sqrt{1 - m^2/\omega^2} $ is the velocity of particle, $l$ is the angular quantum number, $m$ is the mass, $\omega$ is the energy. Since then, this analytical method has
become a standard procedure to calculate absorption cross section.

Meanwhile, there is also the others method to obtain the cross section, for example in 1978 an algebraic method, i.e. power-series approximate expansion, is improved by Sanchezv \cite{Sanchez1,Sanchez2,Sanchez3} in Schwarzschild space. Additionally, Anderson developed a phase-integral method \cite{phase-integral} in 1995. On the aspect of modeling methodology, a wide variety of scattering models have been proposed including fermion emission \cite{fermion,fermion11,fermion22}, charged leptons\cite{charged}, elastic scattering\cite{Sanchez1,Sanchez2,Sanchez3}, Neutrinos\cite{Neutrinos}, D-brane spectroscopy\cite{D-brane,D-brane11} and so on. Attentively, in 1997, Das, Gibbons and Mathur \cite{Das} calculated the low energy absorption cross section of massless scalars and spin 1/2 particles which immerses a general asymptotically flat black hole with arbitrary dimensions. The former cross section is equal to the area of the black hole and the latter one is related to the area of true metric. Unlike this work, a non-asymptotically flat de Sitter universe is considered here.

Recently, more and more attentions paid to high dimensional black hole such as high dimensional Schwarzschild-like black hole with analytical \cite{Schwarzschildlike1,Schwarzschildlike11} and numerical \cite{Schwarzschildlike2,Schwarzschildlike22} methods, high dimensional rotating Kerr-like black holes \cite{Kerr-like,Kerr-like11,Kerr-like22}, high dimensional Schwarzschild-de Sitter-like black hole \cite{sds-like} and so on. Specially, in the last work \cite{sds-like} the absorption and scattering of multi-horizons space was first derived with analytical and numerical techniques by Kanti, Grain and Barrau in the view of high dimensions \footnote{The usual 4D multi-horizons space scattering was firstly studied by G$\ddot{u}$rsel, Sandberg, Novikov, and Starobinsky with charged Reissner Nordstr$\ddot{o}$m black hole \cite{Gursel,Gursel11}.}. It is shown that Hawking radiation can reveal much valuable information such as the space dimensionality, the space curvature, the bulk cosmological constant. Here, we also consider a multi-horizons space but in the view of usual 4 dimensions by the similar method in Ref. \cite{sds-like}.

On the other hand, the accelerating expanded universe is justified by more and more astronomical observations such as Ia Supernovae (SNe Ia) \cite{RuizLapuente,Riess,Branch,Knop,Riess2}, cosmic microwave background (CMB)\cite{Miller,debernardis,Hanany,Halverson} in the Wilkinson Microwave Anisotropy Probe (WMAP). The expanded universe shows us a nonzero cosmological constant $\Lambda$ \cite{supernoa,supernoa11} and the topology structure presents a de Sitter one.
Like the standard null cosmological constant gravity, a lot of literatures focuses on Hawking radiation in an inflationary universe including two-dimensional toy models \cite{Mallett,Mallett11,Mallett22}, radiative tails \cite{Brady}, degenerate special Nariai black hole \cite{Bousso} and so on. In these works, Brady, Chambers, Krivan and Laguna \cite{Brady} demonstrated the existence of exponentially decaying tails at late times outside SdS and Reissner Nordstr$\ddot{o}$m-de Sitter (RNdS) black holes where the S wave mode asymptotes to a nonzero value. In this work, an advanced time $v = t + r_*$ and a retarded time $u = t - r_*$ where $r_*$ is the tortoise coordinate is used. However, there is no work given out the explicit expressions of analytic absorption probability with cosmological constant as far as we know. The reason maybe is the complex ordinary differential equation which could not be solved completely and do not give us the general solution. Considering these situations, we adopt the approximate treatment with low-energy and high-energy limits assumptions and attempt to calculate the absorption and scattering cross sections or greybody factors outside SdS black hole. So the whole problem is divided into two parts: low-energy regime and high-energy regime. In the former case, the similar Unruh's \cite{Unruh} method is used. But we will only use two zones, i.e. near event horizon and cosmological horizon respectively, and match the solution in the intermediate zones. In the latter high-energy case, we use the innermost stable circular orbit to fitting the high energy particles and try to obtain the greybody factor. Another purpose of this paper is to inspect whether the cosmological constant can influence the cross section of usual Schwarzschild black hole.

Finally, we want to clarify two points: one is the units of cosmological constant $\Lambda$ and the other is the polarization of absorption cross section in black hole. For the former case, we use the Planck Units (G=c=$\hbar$=1) to specify $\Lambda$ in this paper. In this units, $\Lambda$ has dimensionless form. Also, in order to simplify the calculation of graybody factor, the mass of black hole $M$ is adopted to unity in this manuscript. Hence, if we convert cosmological constant $\Lambda$ to ordinary units, we can multiply it quantificationally by its conversion factor via $\Lambda \longrightarrow \Lambda \left(GM/c^2\right)$. So in the ordinary units, $\Lambda$ has a dimension of $m^{-2}$. About these specific details, please see the Refs.\cite{Wald,Wald00,Wald11}. For the latter polarization case, incoming scalar particles have no polarization. In fact, the polarization exists in the scattered-out beam for the charged particles outside a black hole, despite the incoming beam is not polarized. To take Reissner-Nordstr$\ddot{o}$m black hole for an example, since the effect of absorption, the non-zero scattering amplitude provides a polarization effect. Hence, the scattered-out beam is polarized, and the polarization vector is perpendicular to the scattering plane. The polarization grade is the function of azimuth angle. About these specific details, please see the Refs.\cite{Wang,Wang11}.

This paper is organized as follows: in section II, we present the
Schwarzschild-de Sitter black hole solution and obtain the general emission equation. In section III, we calculate absorption probabilities for scalar particles emission in low-energy regime. In section IV, we use the geometric optics limit and calculate the greybody factor of scalar particles emission in high-energy regime. Section V is a conclusion. We adopt
the signature $(+, -, -, -)$ and put $\hbar$, $c$, and $G$ equal to
unity.

\section{Schwarzschild-de Sitter black hole and general emission equation}
\begin{figure}
  \includegraphics[width=3.5 in]{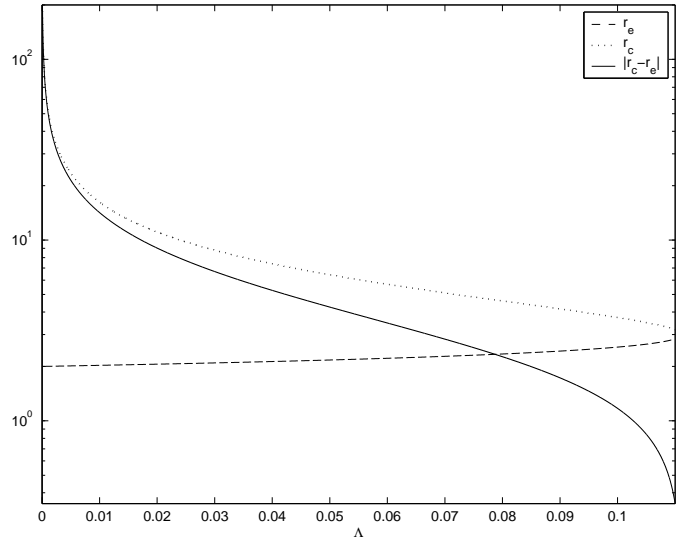}\\
  \caption{event horizon $r_e$ (dashed line), cosmological horizon $r_c$ (dotted line) and interval $|r_c-r_e|$ (solid line) versus cosmological constant $\Lambda$ in the range of $[0,0.11]$ for $M = 1$.}\label{horizons}
\end{figure}

The static spherically symmetric metric of the Schwarzschild-de Sitter
space \cite{Rindler} is
\begin{equation}\label{metric}
    d s^2 = f(r) d t^2 - \frac{1}{f(r)} d r^2 - r^2 \left(d
    \theta^2 +\sin^2\theta d \phi^2\right),
\end{equation}
which is obtained via solving the Einstein gravity field
equations coupling with cosmological constant $\Lambda$,
\begin{equation}\label{EinEQ}
    R_{\mu\nu} - \frac{1}{2} g_{\mu\nu} R + \Lambda g_{\mu\nu} = 8
    \pi G T_{\mu\nu}.
\end{equation}
The central black-hole mass $M$ is contained in metric function $f(r)=1- 2M/r - \Lambda r^2/3$. In this paper, the cosmological constant $\Lambda$ is considered as a free parameter as shown in Refs. \cite{Brevik,Tian,Liu00,Liu0011}. The interesting features for the space metric (\ref{metric}) are its two bounded horizons, one is inner horizon (i.e. black-hole horizon) and the other is outer horizon (i.e. cosmological horizon). Under the limit $\Lambda \longrightarrow 0$,
this metric has exactly the same line-element as the Schwarzschild
space. But in the limit of $M \longrightarrow 0$, it reduces to the
de Sitter one. By using these horizons, the metric function can be rewritten mathematically as a new form,
\begin{equation}
f(r)=\frac{\Lambda}{3r}(r-r_{e})(r_{c}-r)(r-r_{o}). \label{re-f function}%
\end{equation}
The singularity of metric (\ref{metric}) is determined by equation $f(r) =
0$, which is also the null hypersurface condition for static spherically symmetric space. So, these solutions are inner horizon
$r_{e}$ and outer horizon $r_{c}$, as well as a negative solution
$r_{o}=-(r_{e}+r_{c})$. The last one has no physical aspects, and is not considered here. The former two positive solutions $r_e$ and $r_c$ are listed as follows,
\begin{equation}
  r_{c} = \frac{2}{\sqrt{\Lambda}}\cos\eta \ \ \ \  \text{and} \ \ \ \  r_{e} = \frac{2}{\sqrt{\Lambda}}\cos(2\pi/3-\eta), \label{twohorizons}
\end{equation}
where $\eta = 1/3 \arccos(-3M\sqrt{\Lambda}), \left(\eta \in \left[\pi/6 \leq \eta \leq \pi/3\right]\right)$. The real physical solutions are accepted
only if $\Lambda$ satisfies $\Lambda M^2\leq 1/9$ \cite{Liu1}. If the
cosmological constant $\Lambda$ reaches its maximum, the Nariai
black-hole will appear, which has attracted a lot of academic attentions \cite{Nariai,Nariai11,Nariai22}. Here we plot the curves of event horizon $r_e$ and cosmological horizon $r_c$ in Fig.\ref{horizons}, which show clearly that the outer cosmological horizon $r_c$ decreases with increase in cosmological constant $\Lambda$. Else, $r_c$ decreases exponentially quickly when $\Lambda$ is under the order of $10^{-2}$. And the amplitude of variation for $r_c$ is relative slower when $\Lambda$ is larger than $10^{-2}$. But the inner event horizon $r_e$ gently increases oppositely with increase in $\Lambda$. On the other hand, it is known that our world is located between event horizon and cosmological horizon. So we also plot the interval $|r_c-r_e|$ versus cosmological constant $\Lambda$ in the same Fig.\ref{horizons}. It illustrates that with the increasing $\Lambda$, the interval range decreases monotonically. Else, the rate of decreasing is very large both in the primary stage and the terminal stage of dynamic cosmological constant.

Employing the factorized ansatz
\begin{equation}
\Phi=\frac{1}{\sqrt{4\pi\omega}}\frac{1}{r}R_{\omega}(r,t)Y_{lm}(\theta,\phi),\label{wave
function}
\end{equation}
where $Y_{l \omega} (\theta, \phi)$ is the spherical harmonics function, the minimal coupling massless scalar field $\Phi (t, r, \theta, \phi)$ equation
\begin{equation}
\frac{1}{\sqrt{-g}}\frac{\partial}{\partial
x^{\mu}}\left(\sqrt{-g}g^{\mu\nu}\frac{\partial}{\partial
x^{\nu}}\right)\Phi=0,\label{Klein-Gorden equation}
\end{equation}
is broken into spherical harmonic equation (SHE) and the radial radiation equation (RRE). The former SHE is one of the main characteristic for  sphere-symmetric space, and is not shown here. The latter RRE contains much of the information for Hawking radiation, which is shown as,
\begin{equation}
-\frac{1}{f(r)} r^2\frac{\partial^2}{\partial
t^2}\left(\frac{R_{\omega}}{r}\right)+\frac{\partial}{\partial
r}\left(r^2
 f(r)\frac{\partial}{\partial{r}}\left(\frac{R_{\omega}}{r}\right)\right)- l(l+1)\frac{R_{\omega}}{r}=0,\label{radius-t-equation}
\end{equation}
where $R_{\omega}(r, t)$ is the time-dependent radial function.
Radial radiation Eq.(\ref{radius-t-equation}) is the propagated master equation (\ref{radius-t-equation}), which determines the evolution of
evaporating black-hole for scalar particles.

\begin{figure}[h]
\begin{minipage}[t]{0.45\linewidth}
\centering
\includegraphics[width=\textwidth]{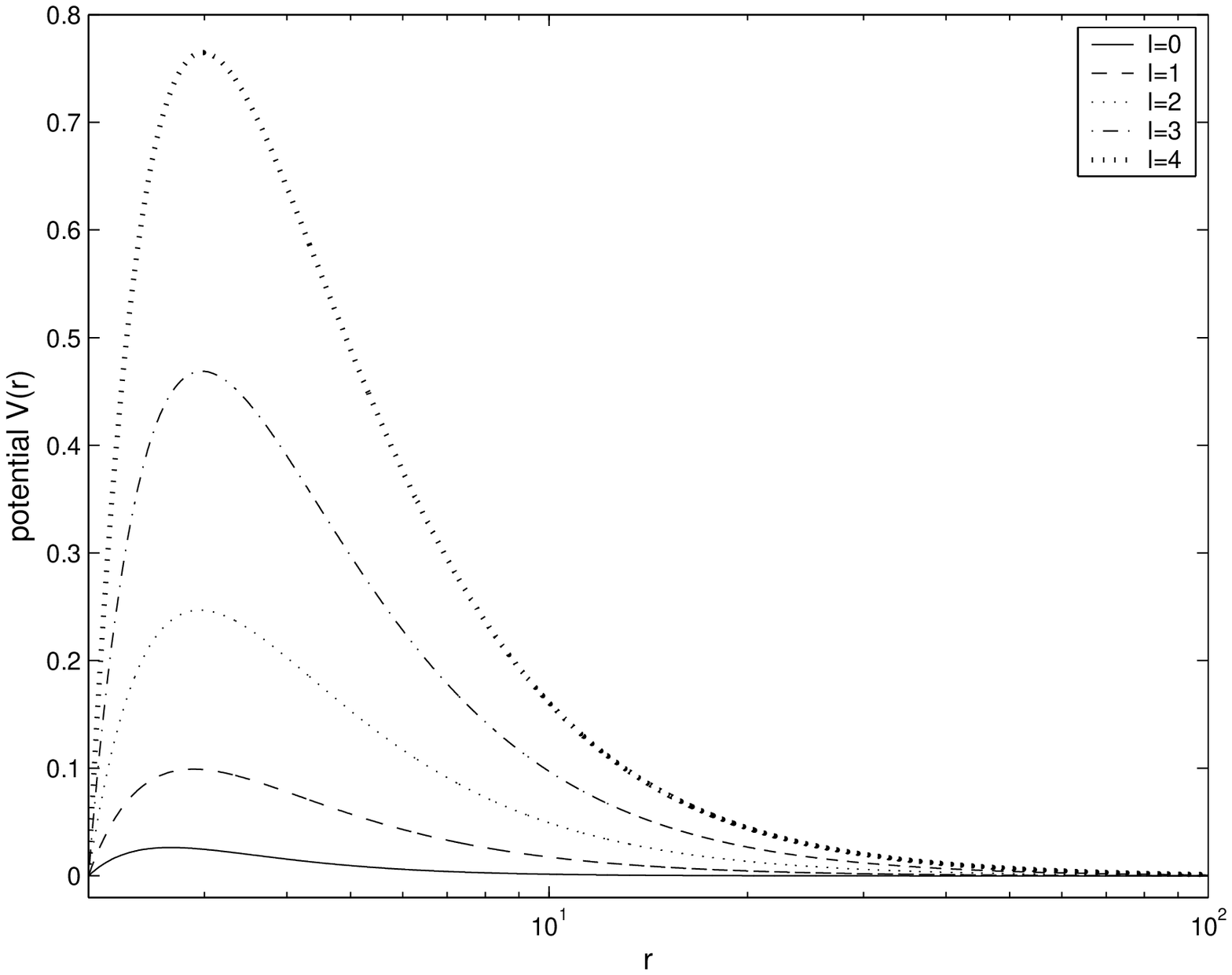}
\caption{potential of Schwarzschild-de Sitter black hole with cosmological constant $\Lambda = 10^{-4}$ for unit $M = 1$ and various angular momentum wave. \label{potential}}
\end{minipage}
\hfill
\begin{minipage}[t]{0.45\linewidth}
\centering
\includegraphics[width=\textwidth]{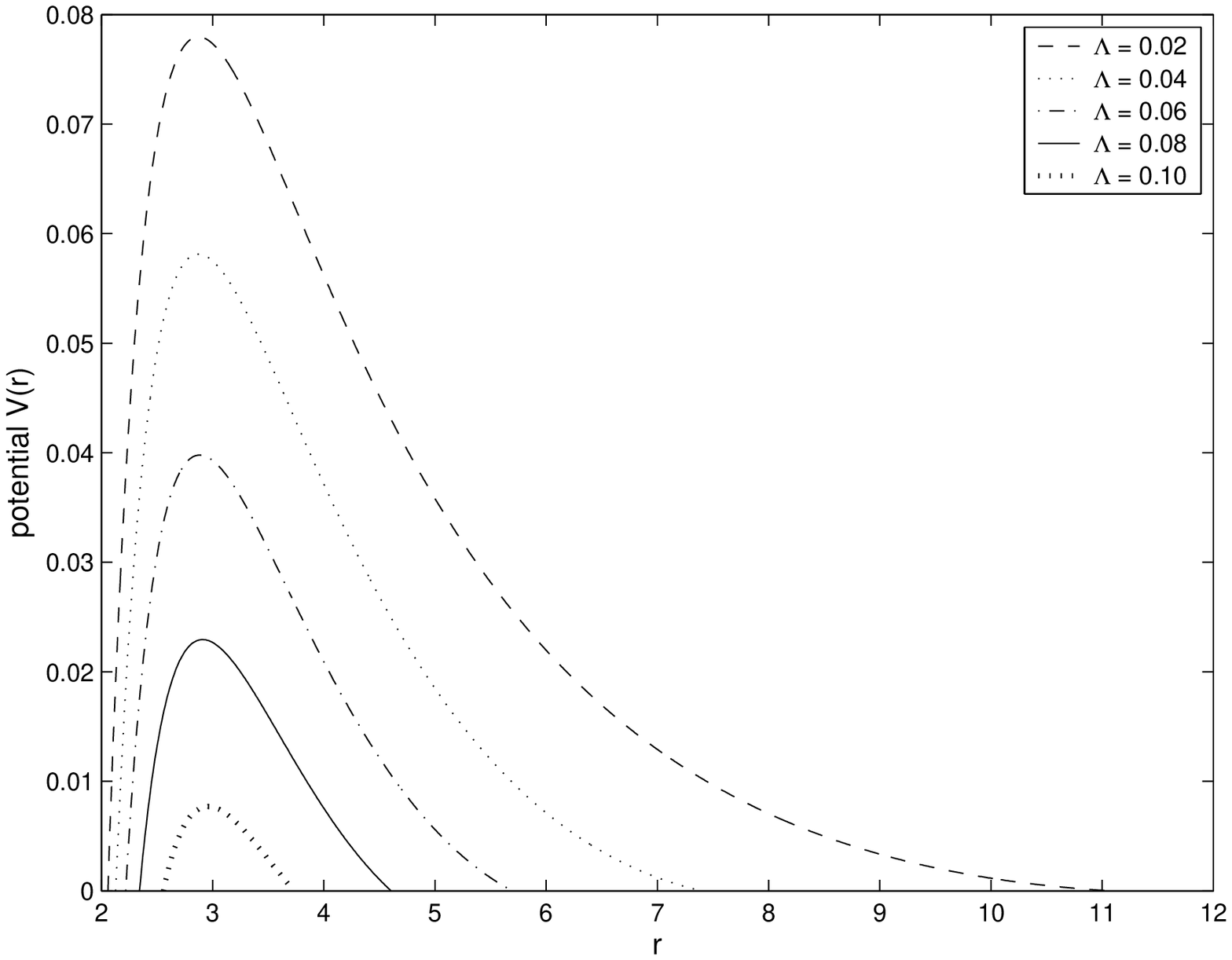}
\caption{potential of Schwarzschild-de Sitter black hole with various cosmological constant for unit $M = 1$ and angular momentum $l =1$ (p wave).\label{potential2}}
\end{minipage}
\end{figure}

Eliminated the time
variable by the Fourier component $e^{-i \omega t}$ via $R_{\omega}(r,t)\rightarrow\Psi_{\omega l}(r) e^{-i\omega t}$,
 Eq.(\ref{radius-t-equation}) can be rewritten as%
 \begin{equation}
 \left[-f(r)\frac{d}{dr}(f(r)\frac{d}{dr})+V(r)\right]\Psi_{\omega
 l}(r)=\omega^2\Psi_{\omega l}(r),\label{radius equ. about r}
 \end{equation}
whose dominion potential is given by%
\begin{equation}
V(r)=f(r)\left[\frac{1}{r}\frac{df(r)}{dr}+\frac{l(l+1)}{r^2}\right].\label{potential-of-r}
\end{equation}
Here in order to inspect the influences of angular momentum and cosmological constant on this potential, we plot Figs.\ref{potential} and \ref{potential2} respectively. From these figures, we can see easily that the potential $V(r)$ depends on parameters $\Lambda$ and $l$ sensitively. It is illustrated that potential $V(r)$ is higher with increase for the angular momentum quantum number $l$ from Fig.\ref{potential}, and it also suggests that for the spherically symmetric s wave, i.e. $l = 0$, its absorption capacity is the strongest than the others. So, the dominant s wave is only considered in this paper. But viewing from Fig.\ref{potential2} the potential becomes lower with increase in cosmological constant.

Now we introduce the tortoise coordinate%
\begin{equation}
r_{*}=\frac{1}{2M}\int\frac{dr}{f(r)}.\label{tortoise }
\end{equation}
The tortoise coordinate can be expressed by surface gravity as
follows,
\begin{equation}\label{tor-grav-sf}
 \nonumber r_{*} = \frac{1}{2M}\bigg{[}\frac{1}{2K_{e}}\ln\left(\frac{r}{r_{e}}-1\right)-\frac{1}{2K_{c}}\ln\left(1-\frac{r}{r_{c}}\right) + \frac{1}{2K_{o}}\ln\left(1-\frac{r}{r_{o}}\right)\bigg{]},
\end{equation}
where the definition of surface gravity is $K_{i}=\frac{1}{2}\left|\frac{df}{dr}\right|_{r=r_i}$.
Using three solutions $r_e$, $r_c$ and $r_o$ to equation $f(r) = 0$,
we can obtain following explicit expressions of corresponding three surface gravities as,
\begin{eqnarray}
  K_{e}=\frac{(r_{c}-r_{e})(r_{e}-r_{o})}{6r_{e}}\Lambda, \\
  K_{c}= \frac{(r_{c}-r_{e})(r_{c}-r_{o})}{6r_{c}}\Lambda,\\
  K_{o}= \frac{(r_{o}-r_{e})(r_{c}-r_{o})}{6r_{o}}\Lambda.
\end{eqnarray}
Through tortoise coordinate transformation (\ref{tortoise }), the radial equation (\ref{radius equ. about r}) can be written in the Regge-Wheeler form
\begin{equation}
\left[-\frac{d^2}{dx^2}+4M^2V(r)\right]\Psi_{\omega
l}(r_{*})=4M^2\omega^2\Psi_{\omega l}(r_{*}),\label{radius-equation}
\end{equation}
which has the form of Schr$\ddot{o}$dinger equation in the quantum
mechanics. The incoming or outgoing particles flowing between inner horizon $r_{e}$
and outer horizon $r_{c}$ are reflected or transmitted by the
potential barrier $V(r)$. It is obvious that the evolution of wave solution
$\Psi_{\omega l}$ of massless scalar field is also determined by
potential $V(r)$.
\section{Absorption probabilities for scalar emission in low-energy regime}
In the above calculation, the radial $r$ is absorbed into wave function $\Psi _{\omega l} (r)$ to get the Regge-Wheeler equation. But in this and next section we need do a reduction run, i.e. using replacement $\psi_{\omega l}(r) = r P_{\omega l} (r)$. So the original radial radiation Eq.(\ref{radius equ. about r}) can be
rewritten to a new form
\begin{equation}\label{solvable1}
    \frac{d}{d r} \left(r^2 f(r) \frac{d P_{\omega l} (r)}{d r}\right) + \left[\frac{\omega^2 r^2}{f(r)} - l(l+1)\right] P_{\omega l} (r) = 0.
\end{equation}
It is well known that, in the low-energy regime $\omega \rightarrow 0$, only the lowest angular momentum will contribute to the cross section \cite{Das}. So, in this paper we consider the dominant scalar decay mode S wave ($l = 0$), which also is illustrated clearly in Fig.\ref{potential}. The analytical solution can be obtained as follows,
\begin{eqnarray}\label{analyticalsolution}
 \nonumber   P_{\omega l}(r) &=& C_2 +C_1\bigg{\{}\frac{\log r}{r_c r_e (r_c + r_e)} + \frac{\log (r-r_c)}{r_c\left(2r_c^2 -r_c r_e -r_e^2\right)}\\
    &+& \frac{\log (r - r_e)}{r_e (-r_c^2-r_c r_e + 2r_e^2)} + \frac{\log (r + r_c +r_e)}{-2r_c^3 -7 r_c^2 r_e -7 r_c r_e^2 -2 r_e^3}
    \bigg{\}}.
\end{eqnarray}
Using the surface gravities to express above solution, we can get
\begin{equation}\label{analytisur}
  P_{\omega l}(r) = C_2 + C_1 \bigg{\{}\frac{\log r}{r_c r_e (r_c +r_e)} -\frac{\Lambda \log (r - r_c)}{6 \kappa _c r_c^2} - \frac{\Lambda \log (r - r_e)}{6 \kappa_e r_e^2} + \frac{\Lambda \log (r + r_e + r_c)}{6 (r_e + r_c)^2 \kappa_o}\bigg{\}}.
\end{equation}
Then we analyse the dynamic behavior of scalar particles propagation near horizons.
According to the forenamed definition of horizons, we know that one feature of a static spherically symmetric space is that the potential vanishes near horizons $r_e$ and $r_c$. So we have the relationship $f(r)|_{(r\rightarrow r_e, r_c)}\longrightarrow 0$ in the nearby regions of horizons $r_e$ and $r_c$ outside SdS black hole immediately. Hence, the Regge-Wheeler equation (\ref{radius-equation}) can be transformed into a standard wave equation form,
\begin{equation}\label{stabdardwaveequa}
\left[\frac{d^2}{dx^2} + 4M^2 \omega^2\right]\Psi_{\omega
l}(r_{*})= 0.
\end{equation}

The general boundary condition of greybody factor is that, near the black hole event horizon the solution is purely ingoing and near the cosmological horizon the solutions are include incoming and outgoing
modes,
\begin{eqnarray}
  \psi_{NE} &=& A_1 e^{-i \omega r_{*}}, \ \ \ \ \text{for}\ \ r \simeq r_e \label{bceh}\\
  \psi_{NC} &=& B_1 e^{- i \omega r_{*}} + B_2 e^{i \omega r_{*}}, \ \ \ \ \text{for}\ \ r \simeq r_c . \label{bcch}
\end{eqnarray}
So near the event horizon $r_c$, the asymptotic solution
(\ref{bceh}) yields a low-energy expansion,
\begin{equation}\label{leeeh}
    \psi^{(e)} \simeq A_1 \bigg{\{}1 - \frac{i\omega}{4 M}\left[\frac{\log (r - r_e)}{\kappa_e} - \frac{\log (r_c - r)}{\kappa_c} + \frac{\log (r - r_o)}{\kappa_o}\right]\bigg{\}}.
\end{equation}
Submitting $r\rightarrow r_e$ into solution $P_{\omega l}$ (\ref{analytisur}), we can get the general solution near event horizon,
\begin{equation}\label{leeeh2}
    \psi^{(e)} = r P_{\omega l}|_{r\rightarrow r_e} = C_2 r_e + C_1 \left[\frac{\log r_e}{r_c r_o} + \frac{\Lambda \log (r - r_e)}{6 r_e \kappa_e} - \frac{r_e \Lambda \log (r_c - r_e)}{6 r_c^2 \kappa_c} + \frac{r_e \Lambda \log (r_e -r_o)}{6 r_o^2 \kappa} \right].
\end{equation}
Comparing Eq.(\ref{leeeh2}) with Eq.(\ref{leeeh}), we can get the expressions of coefficients $C_1$ and $C_2$ near event horizon,
\begin{eqnarray}
  C_1 &=& - A_1 \frac{i 3 \omega r_e}{2 M \Lambda}, \label{C1eh}\\
  C_2 &=& \frac{A_1}{r_e} + \mathcal{O}(\omega).\label{C2eh}
\end{eqnarray}
So near the cosmological horizon, the low-energy expansion of the asymptotic solution (\ref{bcch}) is obtained by using the similar handler routine,
\begin{equation}\label{leeeeee}
    \psi^{c}(r) \simeq (B_1 + B_2) + \frac{i \omega}{4 M} (B_2 - B_1)\left[\frac{\log (r - r_e)}{\kappa_e} -\frac{\log (r_c - r)}{\kappa_c} + \frac{\log (r - r_o)}{\kappa_o}\right].
\end{equation}
Similarly, submitting $r\rightarrow r_c$ into solution $P_{\omega l}$ (\ref{analytisur}), we can get the general solution near cosmological horizon,
\begin{equation}\label{gsnc}
    \psi^{c}(r) = r P_{\omega l}|_{r\rightarrow r_c} = C_2 r_c + C_1 \left[\frac{\log r_c}{r_e r_o} + \frac{r_c \Lambda \log (r_c - r_e)}{6 r_e^2 \kappa_e} -\frac{\Lambda \log (r_c - r)}{6 r_c \kappa_c} + \frac{r_c \lambda \log (r_c - r_o)}{6 r_o^2 \kappa_o}\right].
\end{equation}
Hence, comparing Eq.(\ref{gsnc}) with Eq.(\ref{leeeeee}) we obtain the expressions of coefficients $C_1$ and $C_2$ near cosmological horizon,
\begin{eqnarray}
  C_1 &=& \frac{3}{2} i \omega r_c (B_2 - B_1) \frac{1}{M\Lambda},\label{C1ch} \\
  C_2 &=& \frac{1}{r_c} (B_1 + B_2) + \mathcal{O} (\omega).\label{C2ch}
\end{eqnarray}
Using the solutions near event horizon and cosmological horizon, we then stretch and match them in the intermediate region and obtain the key absorption coefficients. Distinctly, the final relationship of $A_1$, $B_1$, $B_2$ can be obtained according to Eqs.(\ref{C1eh}),(\ref{C1ch}),(\ref{C2eh}),(\ref{C2ch}),
\begin{eqnarray}
  \frac{A_1}{B_1 + B_2} &=& \frac{r_e}{r_c}, \label{frab121}\\
  \frac{A_1}{B_2 - B_1} &=& \frac{r_c}{r_e}.\label{frab122}
\end{eqnarray}
The corresponding reflection and absorption
probabilities are given naturally as follows,
\begin{eqnarray}
  |\mathcal{R}|^2 &=& |\frac{B_2}{B_1}|^2 = \frac{(r_c^2 -r_e^2)^2}{(r_c^2 + r_e^2)^2},\label{reflection} \\
  |\mathcal{A}|^2 &=& 1 - |\mathcal{R}|^2 = \frac{4 r_e^2 r_c^2}{(r_c^2 + r_e^2)^2}.\label{absorption}
\end{eqnarray}
This result is in accordance with that of two-horizon spacetimes \cite{Gursel,Gursel11,Brady}. Here, we present the solution in alternative method. In order to get the variation of absorption probability with cosmological constant, we
use expression Eq.(\ref{twohorizons}) to replace $r_{c}$ and $r_{e}$ in above Eq.(\ref{absorption}). Finally, the analytic expression of absorption probability is obtained as follows
\begin{equation}\label{lowfanaap}
|\mathcal{A}|^2 = \frac{12 + 8\cos 4 \eta_0 - 8\cos 2 \zeta - 8 \sqrt{3} \sin 2\zeta}{18 + 4 \cos 4 \eta_0 - 8\cos 2 \zeta + \cos 4 \zeta - 8 \sqrt{3}\sin 2\zeta -\sqrt{3} \sin 4\zeta},
\end{equation}
where the new variables $\eta_0$, $\zeta$ are defined by $\eta_0 = 1/3 \arccos (3M\sqrt{\Lambda})$, $\zeta = 1/3 \arcsin (3M\sqrt{\Lambda}) $ respectively.
One interesting fact of low energy regime is that this kind of scattering problem varies only with the whole value of $M\sqrt{\Lambda}$ which has a dimension of $GeV^{2}$ coincided with the reciprocal quantity of Newton gravity constant.
\begin{figure}
  \includegraphics[width=3.5 in]{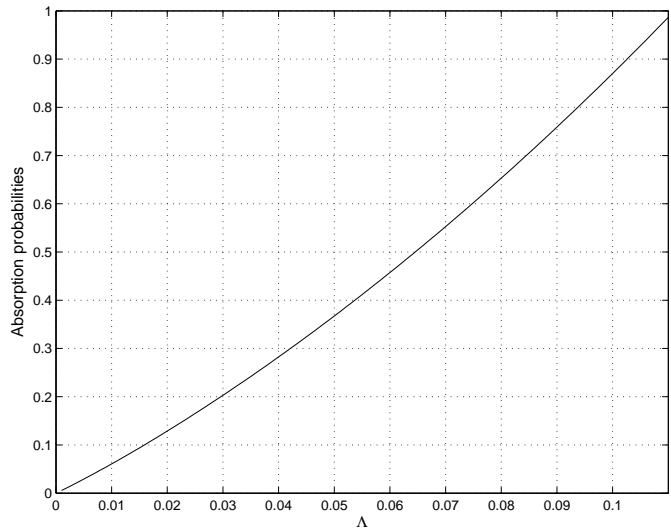}\\
  \caption{absorption probabilities $|\mathcal{A}|^2$ versus cosmological constant $\Lambda$ for unit mass $M = 1$ in low-energy regime.}\label{fig1}
\end{figure}
The form of the absorption probabilities for low-energy regime is shown in Fig. \ref{fig1}. It illustrates that absorption probabilities $|\mathcal{A}|^2$ increases monotonically with cosmological constant. The maximum cosmological constant 0.11 is corresponding maximum probability $|\mathcal{A}|^2_{Max} = 0.986701$. This point also matches the potential $V(r)$ very well (please see Fig. \ref{potential2}). Because that potential becomes lower with increasing cosmological constant, the action of lowering potential on the particles is weak and the absorption probabilities are naturally larger. On the other hand, it is shown that the absorption probability is vanished when cosmological constant disappears. This point is also justified by papers \cite{sds-like,Gursel,Gursel11,Brady}. Referring to the cross section $\sigma_s$ (\ref{crosec}) and transmission coefficients $(T_s)_l$ (\ref{trancoe1}) in 4D Schwarzschild space, we can see that when the particles mass is zero and the energy $\omega$ is more less than the mass of black hole $2M$, the cross section reduces to $\sigma_s \sim 16\pi M^2$. This result shows that, in the low-energy regime the absorption cross section matches accurately with the area of Schwarzschild black hole. This point is expanded to arbitrary dimensions by Das et.al \cite{Das} latterly. From the Eqs. (\ref{crosec}) and (\ref{trancoe1}), we can also see that the transmission coefficients $(T_s)_l$ vanish in the low-energy limit $\omega\longrightarrow 0$ for all values of angular momentum $l$ including s wave, which justify the induction result of zero cosmological constant in SdS black hole. Unlike the singular horizon in Schwarzschild space, there are two different horizons ($r_e$, $r_c$) in Schwarzschild-de Sitter space and the second horizon creates a finite universe. So for the low-energy particles, which has a infinite wavelength, can be localized and has a finite probability to traverse the finite distance between the two horizons \cite{sds-like}.

\section{greybody factor of scalar emission in high-energy regime}
\begin{figure}
  \includegraphics[width=3.5 in]{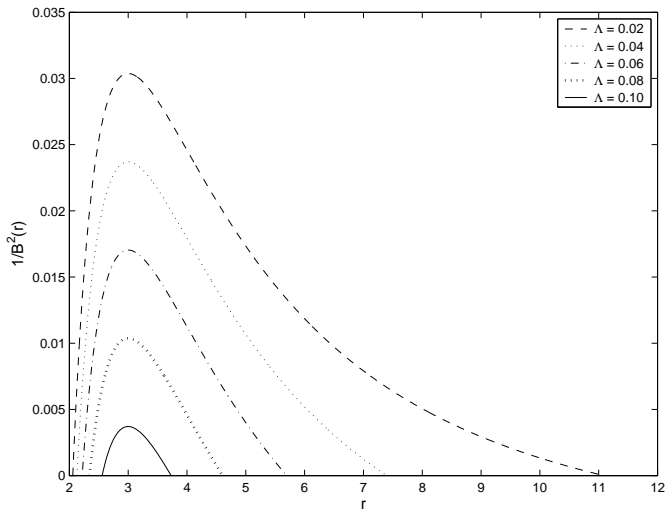}\\
  \caption{ effective potentials $1/B^2 (r)$ of high-energy particles with various cosmological constants and unit mass M = 1.}\label{hheps}
\end{figure}
\begin{figure}
  \includegraphics[width=3.5 in]{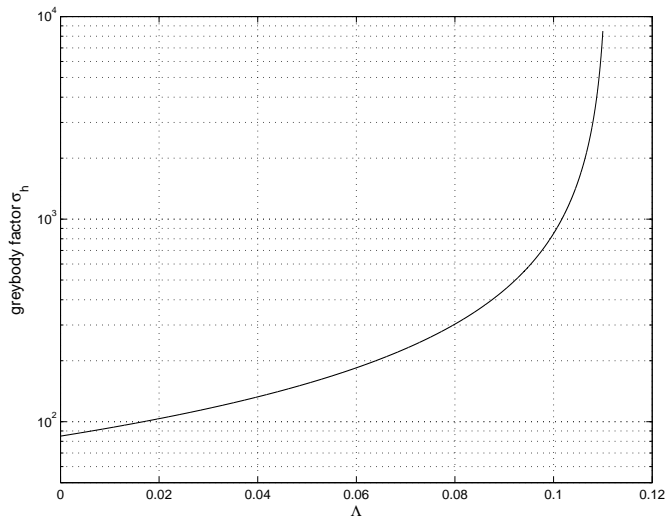}\\
  \caption{greybody factor $\sigma_h$ versus cosmological constant $\Lambda$ for unit mass $M = 1$ in high-energy regime.}\label{greybodyfactorhh}
\end{figure}

When the particles propagate in the high-energy regime, the
scalar greybody factor can be assumed its geometric optics limit value \cite{Sanchez1,Sanchez2,Sanchez3,Emparan,Emparan11,Schwarzschildlike2,Schwarzschildlike22,sds-like,Page1,Page11,Page12}. The momentum of high-energy particle is $p^{\mu} = \frac{d x^{\mu}}{d \zeta}$, where $\zeta$ is an arbitrary scalar parameter for the affine parameter. So the first and second motion equations about $t$ and $\varphi$ are
obtained naturally in the following equations,
\begin{eqnarray}
   && r^2 \frac{d\varphi}{d \zeta} \label{m1} = L,\\
   && (1 - \frac{2M}{r} - \frac{\alpha}{r^{3\omega_q + 1}}) \frac{d t}{d
   \zeta} = \omega\label{m2}.
\end{eqnarray}
And the third motion equation can be given by the normalization relation
of particles $g_{\mu\nu} p^{\mu}p^{\nu}$ = 0,
\begin{equation}\label{m3}
    \left(\frac{d r}{d \zeta}\right)^2 = \omega^2 - \frac{L^2}{r^2}\left(1-\frac{2M}{r}-\frac{\Lambda}{3}r^2\right).
\end{equation}
Then the high-energy particle is subjected Eqs.(\ref{m1}), (\ref{m2}) and (\ref{m3}) near SdS black hole. Combining
Eq. (\ref{m1}) with Eq. (\ref{m3}), the trajectory equation can
be gotten as,
\begin{equation}\label{trajectoryequation}
    \left(\frac{1}{r^2}\frac{d r}{d \varphi}\right)^2 =
    \left(\frac{\omega}{L}\right)^2 - \frac{1}{r^2} \left(1-\frac{2M}{r}-\frac{\Lambda}{3}r^2\right).
\end{equation}
Here, we define two new parameters, one is the impact parameter
$b = L/\omega$, which means the effective sighting range, the other is
the high-energy particle's effective potential $1/B^2(r)$, which is shown in Fig.\ref{hheps}. The parameter $B(r)$ is
\begin{equation}\label{ep}
    B (r) = r \left(1-\frac{2M}{r}-\frac{\Lambda}{3}r^2\right)^{-1/2}.
\end{equation}
According to Eq. (\ref{ep}), particles orbital equation
(\ref{trajectoryequation}) can be rewritten as
\begin{equation}\label{or3}
    \left(\frac{1}{r^2} \frac{d r}{d \varphi}\right)^2 =
    \frac{1}{b^2} - \frac{1}{B^2(r)},
\end{equation}
which is also called Binet equation in the subject of celestial bodies motion orbits. The classically accessible regime is $b_{Max}=B(r)_{Min}$, which means the closest orbit of high-energy particle before absorption by black hole. The
absorption area of the black hole with high energies is
\begin{equation}\label{hgaa}
    \sigma_{h} = \pi b_{Max}^2 = \pi \left(\frac{1}{27M^2} -
    \frac{\Lambda}{3}\right)^{-1},
\end{equation}
which is shown in Fig.\ref{greybodyfactorhh}. Like the regime of low energy, the greybody factor $\sigma_{h}$ is also increasing monotonically with increase in cosmological constant $\Lambda$ in high-energy regime. When the cosmological constant vanishes, the above expression reduces to the
Schwarzschild case $\sigma_{h} = 27\pi M^2 = 27\pi
r_g^2/4$ \cite{Emparan,Emparan11,Schwarzschildlike2,Schwarzschildlike22}.
\section{conclusion}
In this paper, we have studied the scalar particles radiations outside of Schwarzschild-de Sitter black hole in low-energy regime and high-energy regime. Finally, we have successfully obtained the analytic absorption probability $|\mathcal{A}|^2$ in low-energy regime and the greybody factor $\sigma_h$ in high-energy regime, respectively. We find the cosmological constant influences intensely the cross section and greybody not only in the low-energy zone but also in the high-energy zone.

Firstly, we have a look at the case of low-energy limit. It is well known that the cross section of Schwarzschild black hole vanishes in the low-energy and low angular momentum limit \cite{Unruh}. However, in the SdS case the whole space is deformed due to the existence of cosmological constant that changes naturally the scatting of scalar field. For example, the improved potential $V(r)$ (\ref{potential-of-r}), which is illustrated in Figs.(\ref{potential}), (\ref{potential2}), decreases with increasing $\Lambda$. So the absorbent capacity, which expresses mathematically by absorption probabilities $|\mathcal{A}|^2$, enhances for a low potential $V(r)_{low}$. This point also explain essentially the tendency profile of Fig.(\ref{fig1}), which shows absorption probabilities $|\mathcal{A}|^2$ increases with increasing in cosmological constant.

Secondly, we make some discussions on the high-energy limit. Due to the special physical characteristics of  high-energy scalar particles, it is appropriate to assume the geometric optics limit value. Likewise, the greybody factor $\sigma_{h}$ also is under the influence of $\Lambda$. Similarly, the stronger absorption capacity for increase in cosmological constant $\Lambda$ determines the same monotonicity for the greybody factor $\sigma_h$ in low-energy regime in Fig.(\ref{greybodyfactorhh}). The only difference between low-energy and high-energy regimes is that, for the large cosmological constant the $\sigma_{h}$ increases more rapidly.

\acknowledgments
We thank the anonymous referee for helpful suggestions very much. Project is supported by the National Natural Science Foundation (No.10573004), Natural Science Foundation (NSF) (No.10703001), Specialized Research Fund for the Doctoral Program (SRFDP)(No.20070141034) of P.R. China, Henan Educational Committee (No.2010A140015).


\begin{thebibliography}{*}
\bibitem{quanentang}L. Bombelli, R. K. Loul, J. Lee, and R. D. Sorkin, Phys. Rev. D 34 (1986) 373.
\bibitem{quanentang11}D. Kabat, Nucl. Phys. B 453 (1995) 281, hep-th/9503016.
\bibitem{quanentang22} R. M$\ddot{u}$ller and C. Lousto, Phys. Rev. D 52 (1996) 4512, gr-qc/9504049.
\bibitem{entropybekenstein}J. D. Bekenstein, Phys. Rev. D 7 (1973) 2333.
\bibitem{entropybekenstein11}J. D. Bekenstein, gr-qc/9409015.
\bibitem{information}S. W. Hawking, Phys. Rev. D 72 (2005) 084013, hep-th/0507171.
\bibitem{information11} D. N. Page, edited by E. Schmutzer (Friedrich Schiller University, Jena, 1980).
\bibitem{information22} J. Preskill, hep-th/9209058.
\bibitem{information33} D. N. Page, hep-th/9305040.
\bibitem{Hawking}S. W. Hawking, Nature 248 (1974) 30.
\bibitem{Hawking11}S. W. Hawking, Commun. Math. Phys. 43 (1975) 199.
\bibitem{Hawking22}S. W. Hawking, Phys. Rev. D 13 (1976) 191.
\bibitem{Unruh}W. G. Unruh, Phys. Rev. D 14 (1976) 3251.
\bibitem{Sanchez1}N. S$\acute{a}$nchez, Phys. Rev. D 16 (1977) 937.
\bibitem{Sanchez2}N. S$\acute{a}$nchez, Phys. Rev. D 18 (1978) 1798.
\bibitem{Sanchez3}N. S$\acute{a}$nchez, Phys. Rev. D 18 (1978) 1030.
\bibitem{Das}S. R. Das, G. Gibbons, and S. D. Mathur, Phys. Rev. Lett. 78 (1997) 417, hep-th/96090521.
\bibitem{Emparan}R. Emparan, G. T. Horowitz, and R. C. Myers, Phys. Rev. Lett. 85 (2000) 499, hep-th/0003118.
\bibitem{Emparan11}R. Emparan, Nucl. Phys. B 516 (1998) 297, hep-th/97062041.
\bibitem{phase-integral}N. Anderson, Phys. Rev. D 52 (1995) 1808.
\bibitem{fermion} H. T. Cho, A. S. Cornell, J. Doukas and W. Naylor, Phys. Rev. D 77 (2008) 016004, arXiv: 0709.1661.
\bibitem{fermion11} S. Dolan, C. Doran and A. Lasenby, Phys. Rev. D 74 (2006) 064005, gr-qc/0605031.
\bibitem{fermion22} C. Doran, A. Lasenby, S. Dolan and I. Hinder, Phys. Rev. D 71 (2005) 124020, gr-qc/0503019.
\bibitem{charged}D. N. Page, Phys. Rev. D 16 (1977) 2402.
\bibitem{Neutrinos} B. R. Iyer, S. V. Dhurandhar and C. V. Vishveshwara, Phys. Rev. D 25 (1982) 2053.
\bibitem{D-brane}J. Maldacena and A. Strominger, Phys. Rev. D 55 (1997) 861, hep-th/9609026.
\bibitem{D-brane11} S. Das, A. Dasgupta, and T. Sarkar, Phys. Rev. D 55 (1997) 7693, hep-th/9702075.
\bibitem{Schwarzschildlike1}P. Kanti and J. March-Russell, Phys. Rev. D 66 (2002) 024023, hep-ph/0203223.
\bibitem{Schwarzschildlike11}P. Kanti and J. March-Russell, Phys. Rev. D 67 (2003) 104019, hep-ph/0212199.
\bibitem{Schwarzschildlike2}C. M. Harris and P. Kanti, JHEP 10 (2003) 014, hep-ph/0309054.
\bibitem{Schwarzschildlike22} C. M. Harris, hep-ph/0502005.
\bibitem{Kerr-like} D. Ida, K. Oda and S. C. Park, Phys. Rev. D 67 (2003) 064025, hep-th/0212108.
\bibitem{Kerr-like11}D. Ida, K. Oda S. C. Park, Phys.Rev. D 71 (2005) 124039, hep-th/0503052.
\bibitem{Kerr-like22}D. Ida, K. Oda S. C. Park, Phys. Rev. D 73 (2006) 124022, hep-th/0602188.
\bibitem{sds-like}P. Kanti, J. Grain, and A. Barrau, Phys. Rev. D 71 (2005) 104002, hep-th/0501148.
\bibitem{Gursel}Y. G$\ddot{u}$rsel, V. D. Sandberg, I. D. Novikov, and A. A. Starobinsky, Phys. Rev. D 19 (1979) 413.
\bibitem{Gursel11}Y. G$\ddot{u}$rsel, V. D. Sandberg, I. D. Novikov, and A. A. Starobinsky, Phys. Rev. D 20 (1979) 1260.
\bibitem{RuizLapuente}P. Ruiz-Lapuente, A. Burkert, R. Canal, Astrophys. J. 447 (1995) L69, astro-ph/9505090.
\bibitem{Riess}A. G. Riess et al., Astron. J. 116 (1998) 1009, astro-ph/9805201.
\bibitem{Branch}D. Branch, Ann. Rev. Astron. Astrophys. 36 (1998) 17, astro-ph/9801065.
\bibitem{Knop}R.A. Knop et al., Astrophys. J. 598 (2003) 102, astro-ph/0309368.
\bibitem{Riess2} A. G. Riess et al., Astrophys. J. 607 (2004) 665, astro-ph/0402512.
\bibitem{Miller}A. D. Miller et al., Astrophys. J. Lett. 524 (1999) L1, astro-ph/9906421.
\bibitem{debernardis} P. de Bernardis et al., Nature 404 (2000) 955, astro-ph/0004404.
\bibitem{Hanany} S. Hanany et al., Astrophys. J. Lett. 545 (2000) L5, astro-ph/0005123.
\bibitem{Halverson}N. W. Halverson et al., Astrophys. J. 568 (2002) 38, .
\bibitem{supernoa}Supernova Cosmology Project Collaboration, S. Perlmutter et al., Astrophys. J. 517 (1999) 565, astro-ph/9812133.
\bibitem{supernoa11} Supernova Search Team Collaboration, A. G. Riess et al., Astrophys. J. 607 (2004) 665, astro-ph/0402512.
\bibitem{Mallett}R. L. Mallett, Phys. Rev. D 33 (1986) 2201.
\bibitem{Mallett11}P. C.W. Davies, L. H. Ford, and D. N. Page, Phys. Rev. D 34 (1986) 1700.
\bibitem{Mallett22}W. H. Huang, Class. Quantum Grav. 9 (1992) 1199.

\bibitem{Brady}P. R. Brady,  C. M. Chambers,  W. Krivan, and P. Laguna, Phys. Rev. D 55 (1997) 7538, gr-qc/9611056.
\bibitem{Bousso} R. Bousso and S. W. Hawking, Phys. Rev. D 57 (1998) 2436.
\bibitem{Wald}R.M. Wald, {\it General Relativity} (The University of Chicago, Chicago, 1984.
\bibitem{Wald00}I. Brevik1, and B. Simonsen, Gen. Rel. Grav. 33: 1839-1861 (2001).
\bibitem{Wald11}M. Liu, H. Liu, J. Zhang, F. Yu, Chin. Phys. B 17: 1633-1639, (2008),[arXiv:0806.3132].
\bibitem{Wang}Y. J. Wang, {\it Gravity and Cosmology} (Hunan Normal University, Changsha, 2000).
\bibitem{Wang11} C.Q. Liu, Master Dissertation, Hunan Normal University, 2000.

\bibitem{Rindler}W. Rindler, {\it Relativity} (Oxford: Oxford University Press, 2001).
\bibitem{Brevik}I. Brevik and B. Simonsen, Gen. Rel. Grav. 33 (2001) 1839.
\bibitem{Tian}J. X. Tian, Y. X. Gui and G. H. Guo, Gen. Rel.
Grav. 35 (2003) 1473, gr-qc/0304009.
\bibitem{Liu00}M. L. Liu, H. Y. Liu, L. X. Xu and P. S. Wesson, Mod. Phys. Lett. A 21 (2006) 2937, gr-qc/0611137.
\bibitem{Liu0011} M. L. Liu, H. Y. Liu, F. Luo and L. X. Xu, Gen. Rel. Grav. 39 (2007) 1389, arXiv: 0705.2465.
\bibitem{Liu1}H. Y. Liu, Gen. Rel. Grav. 23 (1991) 759.
\bibitem{Nariai}H. Nariai,  Sci. Rep. Tohoku Univ. 34 (1950) 160.
\bibitem{Nariai11}H. Nariai,  Sci. Rep. Tohoku  Univ. 35 (1951) 62.
\bibitem{Nariai22}M. L. Liu, H. Y. Liu, C. X. Wang and Y. L. Ping, Int. J Mod. Phys. A 22 (2007) 4451, arXiv: 0707.0520.
\bibitem{Page1}D. N. Page, Phys. Rev. D 13 (1976) 198.
\bibitem{Page11}D. N. Page, Phys. Rev. D 14 (1976) 3260.
\bibitem{Page12}D. N. Page, Phys. Rev. D 16 (1977) 2402.

\end{thebibliography}
\end{document}